# DROFIT: A LIGHTWEIGHT BAND-FUSED FREQUENCY ATTENTION TOWARD REAL-TIME UAV SPEECH ENHANCEMENT

*Jeongmin Lee[1], Chanhong Jeon[1], Hyungjoo Seo[†2], Taewook Kang[1]*

[1]Sungkyunkwan University, Suwon, Republic of Korea
[2]University of Illinois Urbana-Champaign, Urbana, IL, USA

## ABSTRACT

This paper proposes **DroFiT** (Drone Frequency lightweight Transformer for speech enhancement), a single-microphone speech enhancement network for severe drone self-noise. DroFiT integrates a frequency-wise Transformer with a full/sub-band hybrid encoder–decoder and a TCN back-end for memory-efficient streaming. A learnable skip-and-gate fusion with a combined spectral–temporal loss further refines reconstruction. The model is trained on VoiceBank-DEMAND mixed with recorded drone noise (−5 to −25 dB SNR) and evaluated using standard speech enhancement metrics and computational efficiency. Experimental results show that DroFiT achieves competitive enhancement performance while significantly reducing computational and memory demands, paving the way for real-time processing on resource-constrained UAV platforms. Audio demo samples are available on our demo page*.

## 1. INTRODUCTION

Unmanned aerial vehicles (UAVs) are widely employed for applications including parcel delivery, mountain rescue, and disaster monitoring [1]. Although UAVs primarily rely on vision-based sensing, interest is growing in using acoustic information to enhance situational awareness and interaction [2], [3]. However, wideband, periodic propeller- and motor-induced self-noise severely lowers the captured audio signal-to-noise ratio (SNR), degrading speech and sound processing; thus, effective noise reduction is essential for UAV audition.

Researchers have explored multiple methods to mitigate UAV noise. When multimicrophone arrays are available, conventional multichannel techniques-such as direction-of-arrival (DoA) estimation and beamforming-suppress background noise [4], [5]. Meanwhile, single-channel, deep-learning approaches have been achieving competitive UAV noise suppression with a single microphone, eliminating additional hardware [6], [7], [8]. These learning-based methods directly model the complex relationship between noisy and clean speech in a data-driven manner.

Single-channel deep-learning models generally fall into time-domain and time-frequency-domain approaches. Time-domain methods effectively capture the temporal continuity of speech waveforms and model syllable and phoneme boundaries, thereby improving robustness to nonstationary noise [6], [9]. In contrast, time-frequency-domain approaches exploit temporal and spectral cues, capturing harmonic structures and frequency-dependent noise characteristics-particularly beneficial for UAV noise, which contains strong narrowband harmonic components [10]. Prior works in the time-frequency-domain typically rely on magnitude-only masking, where the spectral magnitude was estimated while reusing the noisy phase [11], [12]. Recently, complex-domain modeling within the time–frequency-domain has gained attention, in which both the real and imaginary components of the short-time Fourier transform (STFT) are jointly estimated [13], [14]. By incorporating phase information alongside magnitude, complex-domain methods achieve more accurate waveform reconstruction than magnitude-only masking, resulting in a more natural sound quality, higher intelligibility, and improved suppression of UAV-specific periodic interference.

Among time-frequency-domain approaches, several architectures have demonstrated strong noise suppression performance in extremely noisy environments (e.g. SNR below -15dB). DCU-Net [14] has emerged as a representative model, achieving high-quality speech enhancement across diverse noise conditions [6]. However, its large parameter count imposes heavy computational and memory demands on low-power, resource-constrained UAV platforms [15]. To address this limitation, the lightweight SMoLnet-T [7] was introduced, significantly reducing parameter count to improve deployment efficiency. Compared to masking-based methods, mapping-based strategies are generally more robust under extremely low-SNR scenarios, since they avoid reliance on the noisy and unreliable phase [8], [16]. Nevertheless, SMoLnet-T relies on chunk-based processing along the time axis as a Transformer requires many input frames to be collected before computation can begin. Such designs introduce latency and necessitate storing entire feature maps, reducing memory-reuse efficiency and increasing peak memory usage. Consequently, SMoLnet-T

[†] Currently at Apple; *https://ml-sp.github.io/DroFiT/

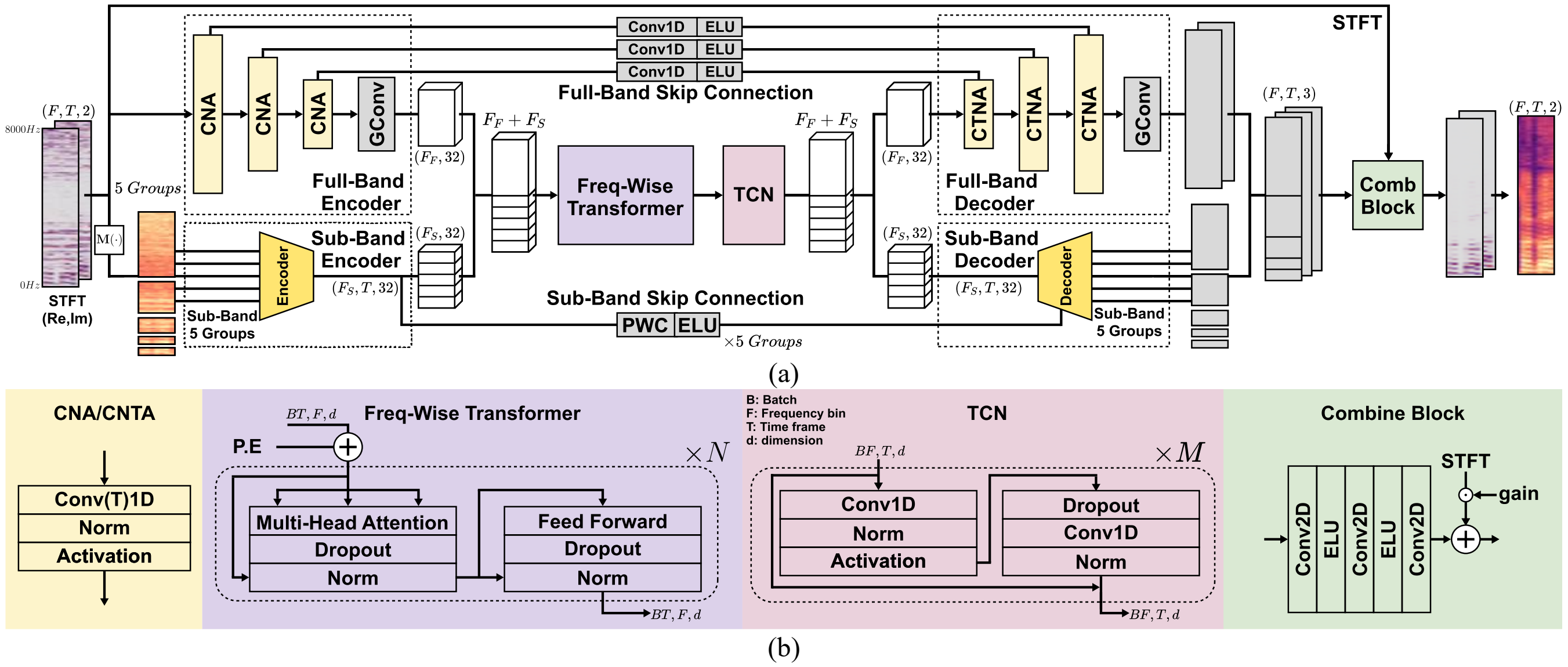

Fig. 1: (a) Overall architecture of the proposed DroFiT, (b) Detailed structure of the submodule.

remains disadvantageous for battery-powered UAVs due to higher power consumption.

To overcome these limitations, we propose DroFiT, which combines efficient temporal and spectral modeling for UAV noise reduction. In the time domain, a temporal convolutional network (TCN) enables real-time streaming with high memory reuse, where only a moderate level of temporal information is sufficient [17], [18]. Unlike time chunk-based models, DroFiT processes audio in small time chunks sequentially, which reduces memory load and improves efficiency. In the frequency domain, a Transformer module captures fine-grained correlations while emphasizing the importance of frequency cues for UAV noise suppression [19]. Moreover, DroFiT employs an encoder to reduce computation and leverages sub-band processing to focus on speech-dominant components [20], [21]. These designs minimize latency and memory usage, making DroFiT suitable for real-time deployment on embedded artificial intelligence (AI) hardware as well as field-programmable gate array (FPGA) and application-specific integrated circuit (ASIC) platforms for battery-powered UAV applications.

## 2. PROPOSED METHOD

In this paper, the input signal is first compressed along the frequency axis and processed by a Transformer to model spectral dependencies. Sub-band and full-band represent-ations are processed in parallel to complement each other instead of using them in series, or cross-connected [22]. The sub-bands provide fine-grained information at each sliced frequency region, especially low-frequency regions containing the speech signal. And the full-band leverages this information to compensate for its inherent limitations, thereby enriching the global spectral context. Fig. 1(a) illustrates the overall architecture of DroFiT, where the input compressed by the encoder is successively processed by N-layers Transformer and M-layers TCN, and the final output is reconstructed by the decoder followed by the combine block.

### 2.1. Full-Band Encoder & Decoder

As illustrated in Fig. 1, the encoder processes the input through a full-band path. The full-band encoder is composed of Conv1D-based CNA blocks, which operate independently along the time axis while compressing features along the frequency axis. A global convolution layer (GConv) is appended to capture long-range spectral dependencies.

The decoder reconstructs the enhanced signal by integrating both encoder representations and the latent module through learnable skip connections, as shown in Fig. 1(a). Unlike fixed skip connections, these projections are parameterized and optimized during training, enabling the network to adaptively balance local sub-band details and global full-band information. This design not only improves the reconstruction quality but also provides greater flexibility in fusing multi-scale spectral features.

### 2.2. Sub-Band Encoder & Decoder

As illustrated in Fig. 1, the sub-band encoder divides the input spectrogram into five groups along the frequency axis, which are transformed into magnitude representations [21]. Specifically, the 513 frequency bins are partitioned into groups of 32-32-64-128-257 bins, respectively, following a Mel-like allocation. Each group is then processed through the path shown in Fig. 2, where lightweight convolutional layers are employed for compression, in contrast to the more complex operations of the full-band encoder.

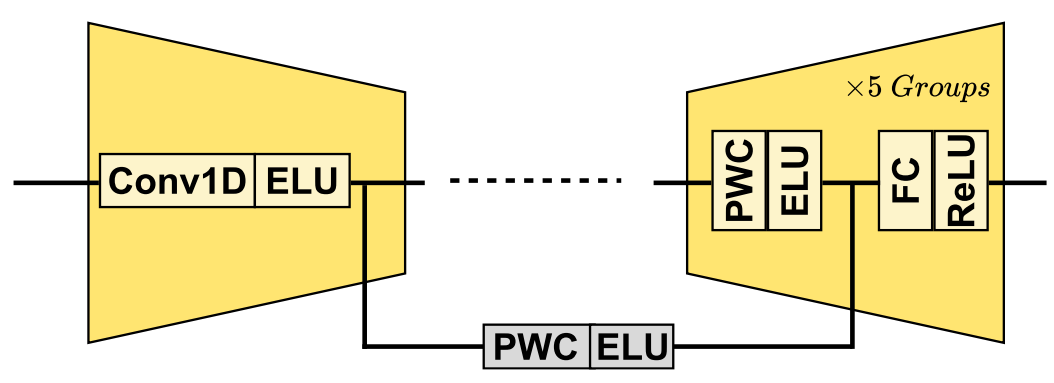

Fig. 2. Architecture of the sub-band encoder–decoder, shown here for one representative group (among the five).

The sub-band decoder follows a structure similar to the full-band decoder, where each group is connected via learnable skip connections. To align the dimensionality of these skip connections with the decoder input, point-wise convolution (PWC) is first applied, followed by a fully connected (FC) layer to restore the original dimensionality.

### 2.3. Frequency-wise Transformer

The core of DroFiT is the Frequency-wise Transformer shown in Fig. 1(b), which applies multi-head self-attention exclusively along the frequency axis while discarding temporal attention, with the full-band and sub-band paths concatenated as its input. Conventional time-frequency attention has a computational complexity of

$$O(F^2T^2d) \tag{1}$$

with $L^2 = F^2T^2$ and $d$ denoting the embedding dimension. In contrast, DroFiT's frequency-only design reduces it to

$$O(F^2Td) \tag{2}$$

DroFiT further reduces complexity in two steps from (2) to (4) by eliminating redundant frequency dimensions. We split the frequency axis into a full-band $F_F$ and sub-band $F_S$ paths with different compression ratios $(k_F, k_S)$, where $F_F = F/k_F$, and $F_S = F/k_S$. The complexity decreases to:

$$O((1/k_F + 1/k_S)^2F^2Td) \tag{3}$$

Second, by restricting the attention span to local windows of size $(w_F, w_S)$, the overall computational complexity is reduced to

$$O((F_Fw_F + 2F_FF_S + F_Sw_S)Td) \tag{4}$$

The model's attention maps are shown in Fig. 3(a), and the effect of the design is further evidenced by the attention map in Fig. 3(b). Rather than permitting global attention across all positions, the method applies local windows. Within each band, self-path attention focuses on capturing fine-grained speech harmonics while suppressing broadly recurring narrowband interference such as UAV noise. Meanwhile, internal-path attention complements missing information between the two bands, thereby preserving a coherent overall representation.

SMoLnet-T has a complexity of $O(FT^2d)$, making streaming infeasible due to quadratic growth on the time axis and the need to reload past key/value states. In contrast, DroFiT replaces temporal attention with a TCN and applies frequency-only local attention, achieving linear time complexity and efficient streaming operation.

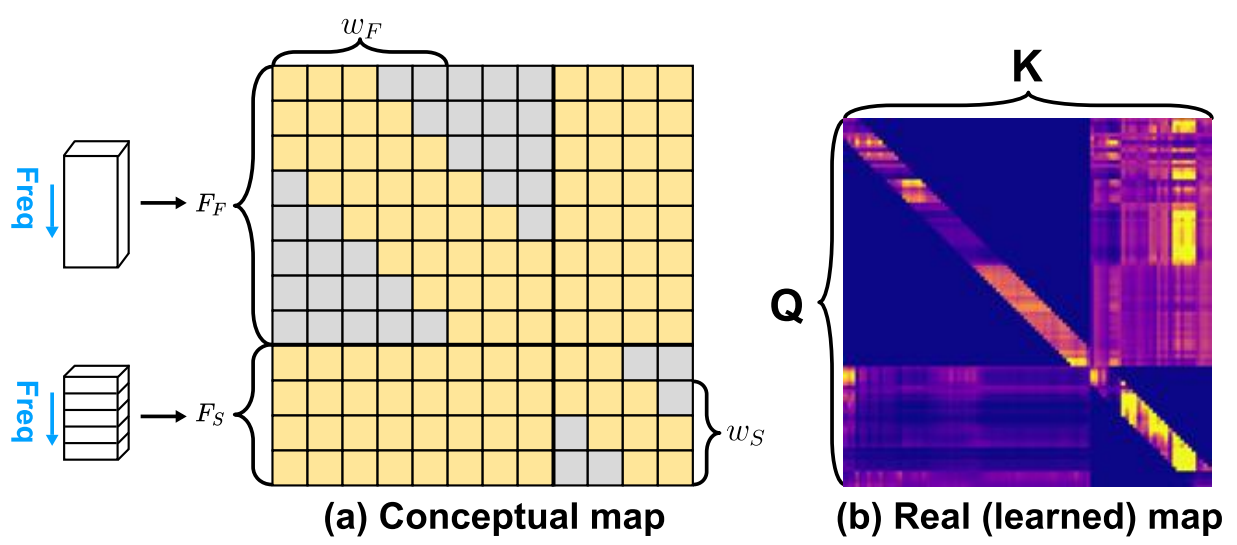

Fig. 3. (a) A conceptual attention map (yellow: attended, gray: masked). (b) An example attention map of DroFiT.

### 2.4. Temporal Convolutional Network (TCN)

After frequency-domain modeling, DroFiT leverages a TCN [18], as shown in Fig. 1(b), to capture temporal dependencies. TCN is composed of Conv1D layers followed by normalization, activation, and dropout. Its receptive field can be flexibly adjusted to either consider only past frames or include both past and future frames, thus enabling efficient modeling over multiple temporal scales.

### 2.5. Output Combination

Finally, DroFiT refines the reconstructed signal through the Combine Block shown in Fig. 1(b). Instead of summing the outputs of the full-band and sub-band paths, the two representations are concatenated and further processed with an additional Conv2D layer and a Learning Gate. The input is organized as three channels (magnitude, real, and imaginary), and the output is reduced to two channels (real and imaginary).

### 2.6. Loss Function Design

To balance spectral consistency with waveform fidelity, the total loss function is defined as a combination of STFT-domain and time-domain losses:

$$Loss = Loss_{STFT} + \alpha Loss_{time}, \tag{6}$$

The STFT-domain loss is a weighted sum of magnitude and complex losses, where $\beta$ controls the weighting.

$$Loss_{STFT} = (1-\beta)Loss_{Mag} + \beta Loss_{Complex}, \tag{7}$$

The magnitude loss minimizes log-spectral differences:

$$Loss_{Mag} = |log_{10}(|\hat{S}_{f,t}|) - log_{10}(|S_{f,t}|)|^2, \tag{8}$$

The complex loss minimizes the difference between estimated and ground-truth STFT spectra, ensuring waveform consistency.

$$Loss_{Complex} = |\hat{S}_{f,t} - S_{f,t}|^2 \tag{9}$$

This design aligns with prior work [23], which emphasized the effectiveness of complex-domain learning for waveform reconstruction.

The time-domain loss is based on the scale-invariant signal-to-distortion ratio (SI-SDR):

$$Loss_{time} = -10\log_{10}\left(\frac{\| s_{target} \|_2^2}{\| r_{noise} \|_2^2 + \varepsilon}\right), \tag{10}$$

where $\boldsymbol{s_{target}} = \frac{\langle \hat{s}, s \rangle}{\|s\|_2^2}\boldsymbol{s}$ is the projection of the estimated signal onto the clean speech, and $\boldsymbol{e_{noise}} = \hat{\boldsymbol{s}} - \boldsymbol{s_{target}}$ is the residual noise. Additionally, auxiliary objectives including time-SI-SDR, cMSE, and LSD are employed to further enhance waveform fidelity and spectral consistency.

## 3. EXPERIMENTS

### 3.1. Dataset Preparation

Drone noise was recorded from a DJI Flip UAV [24] in a hovering state for 9000 seconds and segmented using a stride to create diverse subsets. The recordings were mixed with clean utterances from the VoiceBank-DEMAND [25] corpus (derived from VoiceBank/VCTK [26]), and the test set additionally included one male and one female VCTK speaker not contained in VoiceBank-DEMAND. All utterances were adjusted to 5 seconds. The training set comprised 7200 mixtures at SNRs of −5, −10, −15, −20, and

−25 dB (1440 each), the validation set contained 800 mixtures with the same SNR distribution (160 each), and the test set consisted of 810 mixtures at −5, −10, −15, −20, −25, and −30 dB (135 each). All signals were sampled at 16 kHz, using an STFT with a 1024-point FFT and a 512-sample hop.

### 3.2. Baseline Models

We compare DroFiT to DCU-Net and SMoLnet-T—compact drone-noise suppression models per [6], [7]. DCU-Net employed the wSDR loss function [14] and SMoLnet-T was trained using the same proposed loss function as DroFiT. DroFiT's optimal settings are summarized in Table 2, based on empirical evaluation of different configurations for efficiency-performance trade-offs. These parameters regulate the contributions of the full-band and sub-band paths in DroFiT, balancing performance and efficiency through the compression ratios ($k_F, k_S$) and the window sizes ($w_F, w_S$).

### 3.3. Experimental Settings

The network depth parameters, where N denotes the number of Transformer layers, together with the training configurations, are summarized in Table 1. For consistency, the same definition of N is also applied to SMoLnet-T. In contrast, Table 2 provides detailed hyperparameter settings specific to DroFiT. The Adam optimizer was adopted. All models were trained under the same batch size and optimization settings, and training was performed using double-precision (float64) format [7]. Validation performance determines the number of epochs and hyperparameters to prevent overfitting and ensure fair comparison.

### 3.4. Evaluation Metrics

Model performance was evaluated in terms of both enhancement quality and computational efficiency. The Short-Time Objective Intelligibility (STOI) [27] measures speech intelligibility under noisy conditions, producing a score between 0 and 1. The Extended STOI (ESTOI) [28] refines this measure to better capture intelligibility in non-stationary noise. The Perceptual Evaluation of Speech Quality (PESQ) [29], standardized by ITU-T, predicts subjective listening quality with scores ranging from -0.5 to 4.5. The Scale-Invariant Signal-to-Distortion Ratio (SI-SDR) [30] evaluates waveform reconstruction accuracy in decibels, independent of signal amplitude. Finally, computational efficiency was assessed by reporting multiply–accumulate operations (MACs) and the number of parameters.

TABLE 1. TRAIN PARAMETER SETTING

| Model | Lr | epoch | Loss (α, β) | N | M |
|---|---|---|---|---|---|
| DCU-Net | $10^{-4}$ | 35 | - | - | - |
| SMoLnet-T | | 105 | (0.5, 0.7) | 2 | - |
| DroFiT | | 100 | | 4 | 3 |

TABLE 2. DROFIT MODEL DETAILS

| DroFiT | Full Encoder | Full Decoder | Sub Encoder | Sub Decoder | $w_F, w_S$ | $K_F, K_S$ |
|---|---|---|---|---|---|---|
| Kernel | (6,8,6) | (6,8,6) | (6,8,6) | (6,8,6) | 8, 8 | 64,32 |
| Stride | (2,2,2) | (2,2,2) | (2,2,2) | (2,2,2) | | |

* (sym.: (Layer1-Layer2-Layer3))

## 4. EXPERIMENT RESULTS

We compare the proposed model with speech enhancement networks, namely SMoLnet-T and DCU-Net, which have been applied to drone noise suppression. All models were trained and evaluated under the same dataset and experimental conditions, and performance was assessed using PESQ, STOI, ESTOI, and SI-SDR. The overall results are summarized in Table 3, which also reports the number of parameters and MACs to jointly account for accuracy and computational complexity. Fig. 4 shows the performance across different SNR levels, where the proposed model achieves competitive results comparable to DCU-Net and SMoLnet-T, while consistently outperforming DCU-Net in PESQ, STOI, and ESTOI across all SNR conditions. As observed in Table 3, the proposed model reduces the computational cost by approximately 17.3× and the parameter count by 26.7× compared to DCU-Net. Furthermore, while maintaining similar performance to the drone-targeted SMoLnet-T, it requires fewer parameters and achieves nearly 10× fewer MACs.

## 5. CONCLUSION

We presented DroFiT, a lightweight speech enhancement network designed for UAV self-noise. By combining frequency-wise attention, full/sub-band fusion, and a TCN back-end, the model achieves noise suppression with reduced computational and memory requirements. Trained on mixtures of VoiceBank-DEMAND speech and recorded drone noise at extremely low SNRs, DroFiT delivers competitive quality and intelligibility compared to heavier baselines. The results highlight that restricting attention to the frequency axis and leveraging sub-band processing are key to balancing performance and efficiency. While this work focuses on offline evaluation, lightweight design reduces complexity for real-time, hardware-friendly deployment. Future work will explore broader UAV scenarios and integration with downstream tasks such as ASR and KWS.

TABLE 3. PERFORMANCE & COMPUTATIONAL COMPLEXITY

| Model | PESQ | STOI | ESTOI | SI-SDR | Para. (M) | MAC (G) |
|---|---|---|---|---|---|---|
| SMoLnet-T | 2.433 | **0.669** | **0.440** | **10.315** | 0.187 | 18.64 |
| DCU-Net | 2.433 | 0.639 | 0.421 | 10.129 | 2.808 | 32.23 |
| DroFiT | **2.440** | 0.665 | 0.432 | 9.764 | **0.105** | **1.86** |
| Input Noisy* | 1.666 | 0.561 | 0.261 | -17.10 | - | - |

(Averaged across all test cases) *Input with unsuppressed noise

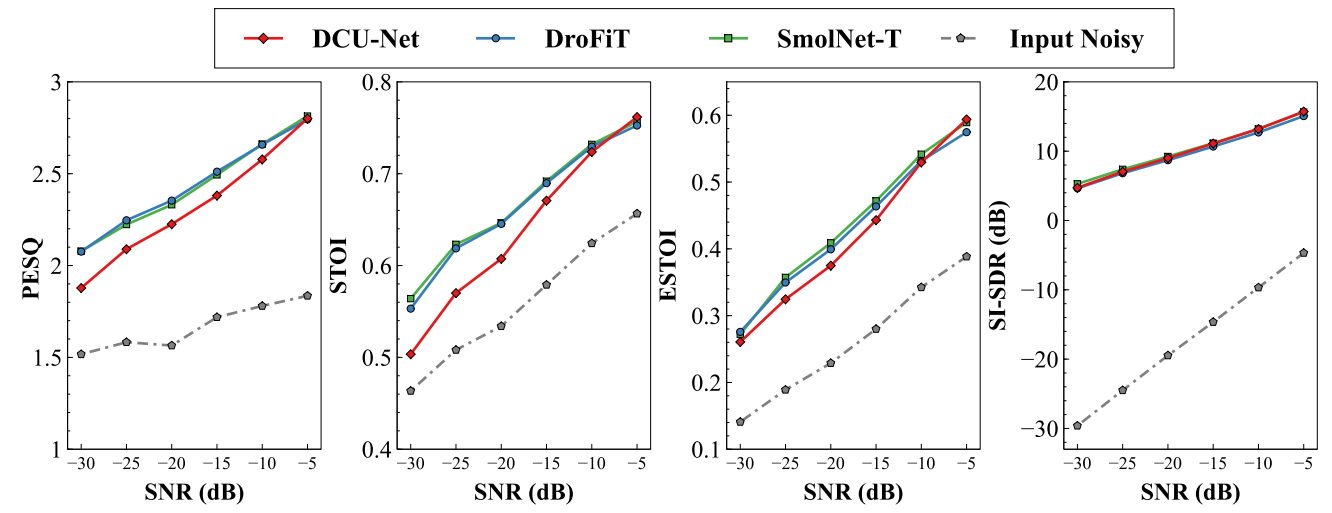

Fig. 4. Test results across SNR levels for each model.

## 6. ACKNOWLEDGEMENTS

This paper was partly supported by the Institute of Information & Communications Technology Planning & Evaluation (IITP) grant funded by the Korea government (MSIT) (No. RS-2025-10692981), and Korea Institute for Advancement of Technology (KIAT), grant funded by the Korea Government (MOTIE). (No. P0023704)

## 7. REFERENCES

[1] M. Guo et al., "Unmanned aerial vehicles in last-mile parcel delivery: A state-of-the-art review," *Drones*, vol. 9, no. 6, Jun., pp. 413, 2025.

[2] Y. Yamazaki, C. Premachandra and C. J. Perea, "Audio-Processing-Based Human Detection at Disaster Sites With Unmanned Aerial Vehicle," in *IEEE Access*, vol. 8, pp. 101398-101405, 2020.

[3] N. Garg and N. Roy, "Enabling self-defense in small drones," *Proc. 21st Int. Workshop on Mobile Computing Systems and Applications (HotMobile)*, Austin, TX, USA, Mar. 2020, pp. 15–20.

[4] L. Wang and A. Cavallaro, "Acoustic Sensing From a Multi-Rotor Drone," *IEEE Sensors Journal*, vol. 18, no. 11, pp. 4570-4582, 1 June1, 2018.

[5] S. Kim, W. Jeong and H. M. Park, "Multi-Channel Speech Enhancement Using Beamforming and Nullforming for Severely Adverse Drone Environment," *IEEE ICASSPW*, Seoul, Korea, Republic of, 2024, pp. 755-759.

[6] D. Mukhutdinov, A. Alex, A. Cavallaro and L. Wang, "Deep Learning Models for Single-Channel Speech Enhancement on Drones," in *IEEE Access*, vol. 11, pp. 22993-23007, 2023.

[7] Z. W. Tan, and A. W. H. Khong, "SMoLnet-T: An efficient complex-spectral mapping speech enhancement approach with frame-wise CNN and spectral combination transformer for drone audition," *Proc. APSIPA ASC*, pp. 1-6, 2024.

[8] Z. W. Tan, A. H. T. Nguyen, and A. W. H. Khong, "An efficient dilated convolutional neural network for UAV noise reduction at low input SNR," *Proc. APSIPA ASC*, pp. 1885–1892, 2019.

[9] S. W. Fu, T. W. Wang, Y. Tsao, X. Lu and H. Kawai, "End-to-End Waveform Utterance Enhancement for Direct Evaluation Metrics Optimization by Fully Convolutional Neural Networks," *IEEE/ACM TASLP*, vol. 26, no. 9, pp. 1570-1584, Sept. 2018.

[10] R. Cabell, R. McSwain, and F. Grosveld, "Measured noise from small unmanned aerial vehicles," *Proc. NOISE-CON*, Providence, RI, USA, Jun. 2016.

[11] A. Narayanan, and D. Wang, "Ideal ratio mask estimation using deep neural networks for robust speech recognition," *Proc. IEEE ICASSP*, pp. 7092–7096, 2013.

[12] Y. Xu, J. Du, L. R. Dai and C. H. Lee, "A Regression Approach to Speech Enhancement Based on Deep Neural Networks," *IEEE/ACM TASLP*, vol. 23, no. 1, pp. 7-19, Jan. 2015.

[13] Y. Hu, et al., "DCCRN: Deep complex convolution recurrent network for phase-aware speech enhancement," *Proc. Interspeech*, pp. 2472–2476, 2020.

[14] H. S. Choi, et al., "Phase-aware speech enhancement with deep complex U-Net," *Proc. Int. Conf. Learning Representations*, 2019.

[15] M. Camurri, D. Palossi, L. Benini, and E. Flamand, "Distilling tiny and ultra-fast deep neural networks for autonomous navigation on nano-UAVs," *IEEE Internet of Things Journal*, 2024.

[16] L. Sun, J. Du, L. R. Dai, and C. H. Lee, "Multiple-target deep learning for LSTM-RNN based speech enhancement," *Proc. Hands-Free Speech Communications and Microphone Arrays*, pp. 136–140, 2017.

[17] T. Peer, and T. Gerkmann, "Phase-aware deep speech enhancement: It's all about the frame length," *JASA Express Letters*, vol. 2, no. 10, Oct. 2022, pp. 104802.

[18] C. Lea, R. Vidal, A. Reiter, and G. D. Hager, "Temporal convolutional networks: A unified approach to action segmentation," *Proc. IEEE CVPRW,* 2017, pp. 156–165.

[19] D. Yin, C. Luo, Z. Xiong, and W. Zeng, "PHASEN: A phase-and-harmonics-aware speech enhancement network," *Proc. AAAI Conf. Artificial Intelligence*, 2020, pp. 9458–9465.

[20] X. Hao, X. Su, R. Horaud, and X. Li, "FullSubNet: A full-band and sub-band fusion model for real-time single-channel speech enhancement," *Proc. IEEE ICASSP*, 2021, pp. 6633–6637.

[21] L. Yang et al., "FSPEN: An ultra-lightweight network for real-time speech enhancement," *Proc. IEEE ICASSP*, Apr. 2024, pp. 10671–10675.

[22] J. Chen, W. Rao, Z. Wang, Z. Wu, Y. Wang, T. Yu, S. Shang, and H. Meng, "Speech Enhancement with Fullband-Subband Cross-Attention Network," *Proc. Interspeech*, 2022, pp. 976–980, doi: 10.21437/Interspeech.2022-10257.

[23] S. Braun and I. Tashev, "A consolidated view of loss functions for supervised deep learning-based speech enhancement," *Proc. International Conference on Telecommunications and Signal Processing*, pp. 72–76, 2021.

[24] DJI, "DJI Flip [Online]," Available: https://www.dji.com/global/flip/

[25] C. Valentini-Botinhao, X. Wang, S. Takaki, and J. Yamagishi, "Noisy speech database for training speech enhancement algorithms and TTS models," dataset, 2017.

[26] J. Yamagishi, C. Veaux, and K. MacDonald, "CSTR VCTK Corpus: English multi-speaker corpus for CSTR voice cloning toolkit," dataset, 2019.

[27] C. H. Taal, R. C. Hendriks, R. Heusdens and J. Jensen, "An Algorithm for Intelligibility Prediction of Time–Frequency Weighted Noisy Speech," *IEEE Transactions on Audio, Speech, and Language Processing*, vol. 19, no. 7, pp. 2125-2136, Sept. 2011.

[28] J. Jensen and C. H. Taal, "An algorithm for predicting the intelligibility of speech masked by modulated noise maskers," IEEE/ACM Trans. Audio, Speech, Language Process., vol. 24, no. 11, pp. 2009–2022, 2016.

[29] A. W. Rix, J. G. Beerends, M. P. Hollier, and A. P. Hekstra, "Perceptual evaluation of speech quality (PESQ) – a new method for speech quality assessment of telephone networks and codecs," *Proc. IEEE ICASSP*, vol. 2, pp. 749–752, 2001.

[30] J. L. Roux, S. Wisdom, H. Erdogan, and J. R. Hershey, "SDR – half-baked or well done?" *Proc. IEEE ICASSP* pp. 626–630, 2019.